# Magnetic properties of the quasi-one-dimensional $S = 1$ spin chain antiferromagnet BaNiTe$_2$O$_7$

Xiyu Chen,[1,2,#] Yiming Gao,[2,#] Meifeng Liu,[1,*] Tao Zou,[3,*] V. Ovidiu Garlea,[4] Clarina dela Cruz,[4] Zhen Liu,[1] Wenjing Niu,[1] Leili Tan,[1] Guanzhong Zhou,[5] Fei Liu,[1] Shuhan Zheng,[1] Zhen Ma,[1] Xiuzhang Wang,[1] Hong Li,[1] Shuai Dong,[2,*] Jun-Ming Liu[1,5]

[1]*Institute for Advanced Materials, College of Physics and Electronic Science, Hubei Normal University, Huangshi 435002, China*

[2]*Key Laboratory of Quantum Materials and Devices of Ministry of Education, School of Physics, Southeast University, Nanjing 211189, China*

[3]*Collaborative Innovation Center of Light Manipulations and Applications, Shangdong Normal University, Jinan 250358, China*

[4]*Neutron Scattering Division, Oak Ridge National Laboratory, Oak Ridge, Tennessee 37831, USA*

[5]*Laboratory of Solid State Microstructures, Nanjing University, Nanjing 210093, China*

**ABSTRACT**

We report a quasi-one-dimensional $S = 1$ spin chain compound BaNiTe$_2$O$_7$. This magnetic system has been investigated by magnetic susceptibility, specific heat, and neutron powder diffraction. These results indicate that BaNiTe$_2$O$_7$ develops a short-range magnetic correlation around $T \sim 22$ K. With further cooling, an antiferromagnetic phase transition is observed at $T_N \sim 5.4$ K. Neutron powder diffraction revealed antiferromagnetic noncollinear order with a commensurate propagation vector **k** = (1/2, 1, 0). The refined magnetic moment size of Ni$^{2+}$ at 1.5 K is 1.84$\mu_B$, and its noncollinear spin texture is confirmed by first-principles calculations. Inelastic neutron-scattering results and density functional theory calculations confirmed the quasi-one-dimensional nature of the spin systems.

[#] The two authors contribute equally to this work.

[*] Corresponding authors: lmfeng1107@hbnu.edu.cn; taozoucn@gmail.com; sdong@seu.edu.cn

## I. INTRODUCTION

Frustrated and low-dimensional magnetic interactions can induce exotic magnetic ground states and novel quantum phenomena in the quasi-one-dimensional magnets, and thus they have been extensively explored in recent years [1-7]. Theoretical study predicted that one-dimensional isotropic magnets with finite-range exchange interaction can not stabilize long-range magnetic ordering at nonzero temperatures [8]. In real materials, interchain interactions invariably survive. Depending on the magnitude of the interchain interactions, quasi-one-dimensional magnets would exhibit long-range magnetic ordering or magnetic field-induced long-range magnetic ordering [9-12]. In addition, the very rich physics has been reported in such quasi-one-dimensional magnets, such as Bose-Einstein condensation [13], spin-induced multiferroicity [14-18], quantum spin liquid [1], and topological quantum phase transition [19].

Since Haldane predicted that integer spin chains have a gapped excitation spectrum between ground state and excited state, which is different from half-integer spin chains, the integer spin systems received extensive investigations [20,21]. Experimental evidence of a Haldane gap has been found in many quasi-one-dimensional $S = 1$ spin chain compounds [9,22-26]. Furthermore, the weak interchain interactions and anisotropy would also affect the magnetic ground state [10,11,23,26-28]. Long-range magnetic ordering has been observed in $CsNiCl_3$ and $NiTe_2O_5$ [11,29], while $PbNi_2V_2O_8$ and $SrNi_2V_2O_8$ were reported to exhibit a nonmagnetic singlet ground state [9,10,26]. Interestingly, a magnetic field-induced magnetic ordering state was found in $PbNi_2V_2O_8$ and $SrNi_2V_2O_8$ [9,10]. The former was believed to be a field-induced Bose-Einstein condensation, while the latter was attributed to a field-induced three-dimensional Néel antiferromagnetic ordering and anisotropy plays an important role [9,10]. Such rich physics attracts extensive attention in $S = 1$ quasi-one-dimensional magnets.

Recently, a new noncentrosymmetric family of $BaMTe_2O_7$ ($M = Mg^{2+}$, $Co^{2+}$, $Cu^{2+}$, and $Zn^{2+}$) compounds was intensively studied, where $M^{2+}$ ions form quasi-one-dimensional zigzag chains and the tellurium element has two different valence states, $Te1^{6+}$ and $Te2^{4+}$ [30-32]. Earlier investigations have shown that $BaMTe_2O_7$ ($M = Mg^{2+}$, $Cu^{2+}$, and $Zn^{2+}$) belongs to polar materials, but it is nonferroelectric because the polarization cannot be reversed by the external electric field [30,31]. In addition, the magnetic susceptibility of $BaCuTe_2O_7$ shows a broad peak around 71 K; no long-range magnetic ordering was found down to 5 K [30]. Very recently, the structure and magnetic properties of the spin chain $BaCoTe_2O_7$ were investigated [32]. These results show that $BaCoTe_2O_7$ has a long-range magnetic ordering at $T_N$ = 6.2 K, and a short-range antiferromagnetic spin correlation was observed around 20 K [32]. Furthermore, the neutron-diffraction data suggest $BaCoTe_2O_7$ exhibits a canted ↑↑↓↓ magnetic structure along the chain below $T_N$ = 6.2 K [32]. Thus, the spin-induced ferroelectricity was claimed to possibly exist in $BaCoTe_2O_7$ [32]. This family exhibits very interesting physical phenomena. However, $BaNiTe_2O_7$ still has not been investigated.

In this paper, we firstly report the structure and magnetism of $BaNiTe_2O_7$ by x-ray diffraction, magnetic susceptibility, specific heat, neutron powder diffraction (NPD), and inelastic neutron-scattering techniques. Our results show that $BaNiTe_2O_7$ crystallizes in a non-centrosymmetric space group *Ama*2 (No. 40) and $Ni^{2+}$ ions form spin chains running along the *a*-axis. A commensurate antiferromagnetic ordering with a propagation vector **k** = (1/2, 1, 0) is found below $T_N$ ~ 5.4 K, and the inelastic spectrum indicates a strong one-dimensional character of spin excitations. These observations are further confirmed by first-principles calculations.

## II. EXPERIMENT AND CALCULATION

Polycrystalline samples of $BaNiTe_2O_7$ were synthesized by the conventional solid-state reaction method. The stoichiometric mixture of $BaCO_3$, $NiO$, and $TeO_2$ was ground and fired at 550 °C and 650 °C for 24 hours in air, respectively. The obtained powders were reground and pelleted. The resultant pellets were sintered at 750 °C for 24 h in air. Finally, x-ray diffraction (XRD) was performed using the SmartLab Se x-ray diffractometer (Cu *Kα* radiation) at room temperature (*T*). The diffraction data were further analyzed by the Rietveld refinement based on GSAS package.[33,34] The sample masses used to measure magnetism, specific heat and neutron diffraction are 11.9 mg, 4.3 mg, and 1.5 g, respectively. The temperature-dependent magnetic susceptibility *χ*(*T*) and specific heat $C_p(T)$ was measured using physical property measurement system (PPMS DynaCool-9, Quantum Design). Neutron powder diffraction data were collected with neutron wavelength *λ* = 2.41 and 1.54 Å at the HB2A beamline located at the High Flux Isotope Reactor, Oak Ridge National Laboratory. The FullPROF package was used to refine both the nuclear and magnetic structures [35]. Inelastic neutron-scattering (INS) measurements were performed using the HYSPEC time-of-flight spectrometer at the Spallation Neutron Source. Measurements were taken using an incident energy of $E_i$ = 13 meV and a Fermi chopper speed of 360 Hz. Analysis of the magnet excitations was performed with linear spin-wave theory using the SPINW program [36].

Density functional theory (DFT) calculations are performed using the Vienna *ab initio* Simulation Package [37]. The electronic interactions are described by projector-augmented-wave pseudopotentials with semicore states treated as valence states [38]. To precisely describe the crystal structure, the generalized gradient approximation with Perdew-Burke-Ernzerhof functional modified for solids (PBEsol) parametrization is adopted [39]. To count the correlation effect, the Hubbard $U_{eff} = U-J = 4$ eV is applied to Ni's 3$d$ orbitals [40]. The plane-wave cutoff energy is fixed as 500 eV. The $k$-point grids of 6 × 2 × 4 are adopted. The convergent criterion for the energy was set to $10^{-5}$ eV, and that of the Hellman-Feynman forces during the structural relaxation is 0.01 eV/Å. In addition, spin-orbit coupling (SOC) is taken into consideration.

## III. RESULTS AND DISCUSSION

Figure 1(a) displays the crystal structure of BaNiTe$_2$O$_7$ with a non-centrosymmetric space group *Ama*2. The NiO$_5$ square pyramids form quasi-one-dimensional zigzag spin chains along the *a* direction. The intrachain interactions can be defined by two distinct exchange parameters ($J_1$, $J_2$) corresponding to different Ni$^{2+}$-Ni$^{2+}$ distances ($d_{J1}$ = 4.6 Å, $d_{J2}$ = 5.56 Å). Both couplings involve superexchange paths mediated by two distinct tellurium atoms, possessing two different valence states, Te1$^{6+}$ and Te2$^{4+}$, which exist in Te1O$_6$ octahedral and Te2O$_4$ polyhedral coordination environments. The Ni-O-Te-O-Ni bond angles corresponding to $J_1$ and $J_2$ are very different, thus the two exchanges are expected to have different strengths. In addition, the interchain interactions can be described as $J_3$ with Ni$^{2+}$-Ni$^{2+}$ distance of 5.94 Å. First-principles calculations, described below, were performed to estimate the $J_1/J_2/J_3$ values. The Rietveld refinement of the XRD pattern for BaNiTe$_2$O$_7$ is shown in Fig. 1(b). All typical diffraction peaks can be indexed with the space group *Ama*2, and the refined lattice parameters are $a$ = 5.5649(63)

Å, $b$ = 15.1227(84) Å, $c$ = 7.2685(93) Å, and $\alpha = \beta = \gamma = 90°$. Table I displays the detailed lattice parameters of BaNiTe$_2$O$_7$.

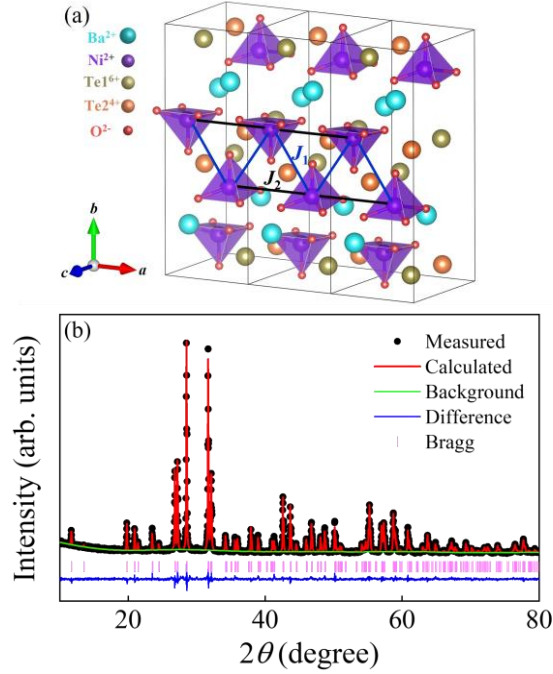

FIG. 1. (a) The crystal structure of BaNiTe$_2$O$_7$. (b) Rietveld refined powder XRD spectrum of BaNiTe$_2$O$_7$ at room temperature.

TABLE I. Refined lattice parameters of BaNiTe$_2$O$_7$ from XRD at room temperature.

| Atom (site) | $x$ | $y$ | $z$ | Occ. | $U_{iso}$ |
|---|---|---|---|---|---|
| Ba1(4b) | 0.250 | 0.2082(4) | 0.9522(8) | 1.000 | 0.0047(0) |
| Ni1(4b) | 0.250 | 0.1256(5) | 0.4676(1) | 1.000 | 0.0175(6) |
| Te1(4b) | 0.250 | 0.9248(1) | 0.7067(6) | 1.000 | 0.0077(2) |
| Te2(4b) | 0.750 | 0.0736(5) | 0.1904(9) | 1.000 | 0.0042(3) |
| O1(4b) | 0.750 | 0.1460(1) | -0.0418(9) | 1.000 | 0.0053(9) |
| O2(8c) | 0.0345(3) | 0.1481(2) | 0.2560(9) | 1.000 | 0.0347(6) |
| O3(4b) | 0.250 | 1.0065(9) | 0.4514(7) | 1.000 | 0.0728(0) |
| O4(8c) | -0.0132(6) | 0.8427(0) | 0.6297(9) | 1.000 | 0.0374(9) |
| O5(4a) | 0.000 | 0.000 | 0.8472(0) | 1.000 | 0.0235(2) |

Space group: $Ama2$, $R_p$ = 5.20%, $R_{wp}$ = 6.74%, $\chi^2$ = 1.79

Figure 2(a) presents the temperature-dependent magnetic susceptibility $\chi(T)$ of BaNiTe$_2$O$_7$ under magnetic field $\mu_0H = 0.1$ T. Both $\chi(T)$ curves acquired after zero-field cooling and field cooling processes almost overlap. Besides, a broad peak shows up around 22 K, which indicates the occurrence of a short-range magnetic correlation in low-dimensional magnets corresponding to the quasi-one-dimensional Ni$^{2+}$-based zigzag spin chains. The inverse magnetic susceptibility $1/\chi(T)$ can be well fitted by the linear Curie-Weiss law $\chi = C/(T - \theta_{CW})$ above $T = 150$ K, where $C$ is Curie-Weiss constant and $\theta_{CW}$ is Curie-Weiss temperature. The fitted parameters are $C \sim 1.00$ emu K/mol and $\theta_{CW} \sim -19.61$ K. The effective magnetic moment $\mu_{eff} \sim 2.83$ $\mu_B$/Ni$^{2+}$ obtained from the fitting is consistent with the expected spin-only value of $\mu_{eff} = g\sqrt{S(S+1)} \sim 2.83 \mu_B$ for Ni$^{2+}$ ($S = 1$, $g = 2$). The negative $\theta_{CW}$ indicates that antiferromagnetic (AFM) interactions dominate in BaNiTe$_2$O$_7$.

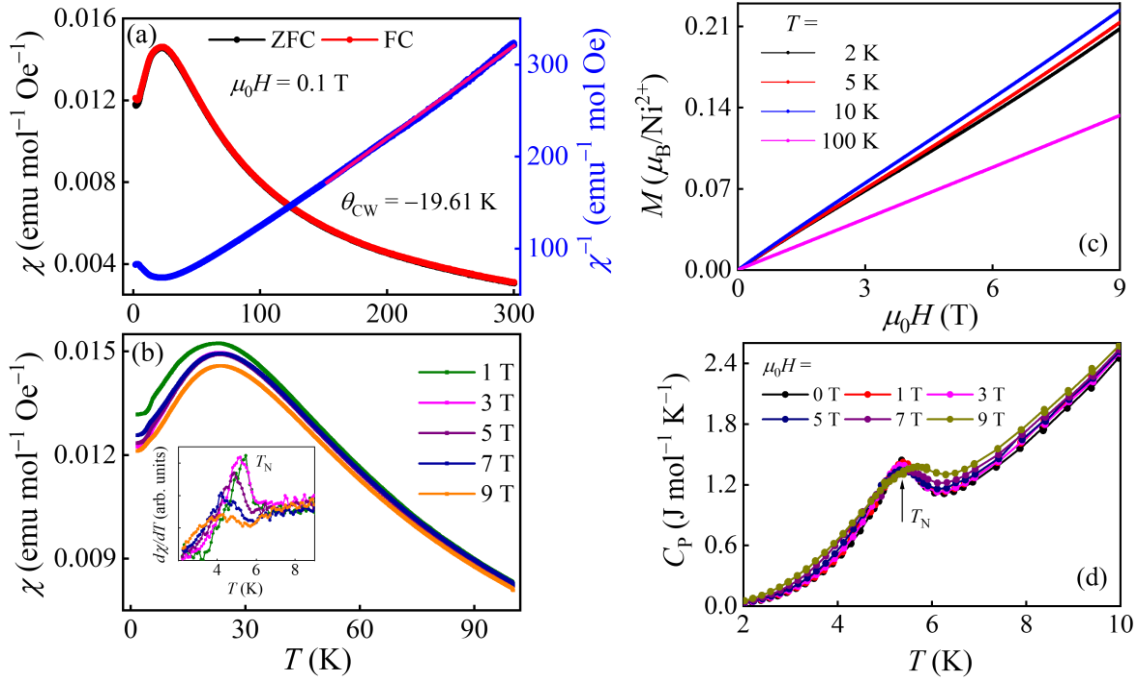

FIG. 2. (a) Temperature-dependent magnetic susceptibility $\chi(T)$ of BaNiTe$_2$O$_7$ measured under $\mu_0H = 0.1$ T (red). The inverse magnetic susceptibility and corresponding linear Curie-Weiss fit (blue). (b) $\chi(T)$ measured under various magnetic fields and the inset displays the derivative $d\chi(T)/dT$ around 5 K. (c) Magnetic field-dependent magnetization $M(\mu_0H)$ of BaNiTe$_2$O$_7$ measured at $T = 2, 5, 10$ and 100 K. (d) Temperature-dependent specific heat $C_P(T)$ under zero and various magnetic fields.

The $\chi(T)$ curves of BaNiTe$_2$O$_7$ were also measured under various magnetic fields ranging from 1 to 9 T, as presented in Fig. 2(b). All curves exhibit a broad peak around 22 K, suggesting the occurrence of short-range magnetic order. The derivative $d\chi(T)/dT$ shows a sharp peak at $T_N$ ~ 5.4 K, indicating that BaNiTe$_2$O$_7$ undergoes an AFM phase transition. In addition, one can clearly see that the AFM ordering of BaNiTe$_2$O$_7$ is suppressed by a field above 7 T. Moreover, the frustration factor $f = |\theta_{CW}|/T_N$ ~ 3.63, demonstrates the existence of moderate magnetic frustration in BaNiTe$_2$O$_7$. Compared with other Ni-based quasi-one-dimensional magnets PbNi$_2$V$_2$O$_8$ and SrNi$_2$V$_2$O$_8$, BaNiTe$_2$O$_7$ potentially has a relatively larger interchain interaction which leads to the AFM phase transition under zero magnetic field [9,10]. To investigate the effect of the magnetic field on magnetism, the field-dependent magnetization $M(\mu_0 H)$ was measured at various temperatures, as shown in Fig. 2(c). The $M(\mu_0 H)$ curves exhibit linear behavior, and no magnetic field-induced metamagnetic transitions are observed under the magnetic fields up to 9 T. In a recent study, a field-induced spin-flop transition is found in BaCoTe$_2$O$_7$ single crystals [32]. Therefore, it is worthwhile to grow BaNiTe$_2$O$_7$ single crystals and further investigate theirs magnetic properties in the future.

The specific heat $C_P(T)$ of BaNiTe$_2$O$_7$ as a function of temperature was also measured to further clarify the AFM phase transition, as presented in Fig. 2(d). A pronounced anomalous peak is observed at $T_N$ ~ 5.4 K, confirming the AFM phase transition, consistent with the magnetic susceptibility $\chi(T)$. The peak position slightly shifts to lower $T$ and the magnitude gradually decreases with increasing magnetic fields. Interestingly, this anomalous peak becomes slightly broader and shifts towards higher temperatures when magnetic fields $\mu_0 H$ is above 5 T. This unusual behavior probably originates from the effect of the magnetic field on the frustrated interchain interactions of BaNiTe$_2$O$_7$, giving rise to the instability of magnetic ordering.

To investigate the AFM ground state of BaNiTe$_2$O$_7$, NPD data were firstly collected at 20 K, which is above the $T_N$ ~ 5.4 K. The neutron wavelength 1.54 Å was used to better characterize the nuclear lattice parameters. Both the experimental and refined data are plotted in Fig. 3(a). All the Bragg reflections can be well indexed with the space group *Ama*2, consistent with the XRD shown in Fig. 1(b). The obtained lattice parameters are $a$ = 5.56240(11) Å, $b$ =

15.09720(33) Å, $c$ = 7.26325(15) Å, and $α = β = γ = 90°$. To better understand the magnetic structure of BaNiTe$_2$O$_7$, NPD data was subsequently collected at 1.5 K with neutron wavelength 2.41 Å. Both the experimental and refined data are presented in Fig. 3(b). A series of magnetic Bragg peaks appear as indicated by the second row of olive vertical bars. Note that no obvious lattice parameter changes were detected. The magnetic peak Bragg positions can be well indexed with a commensurate propagation vector **k** = (1/2, 1, 0).

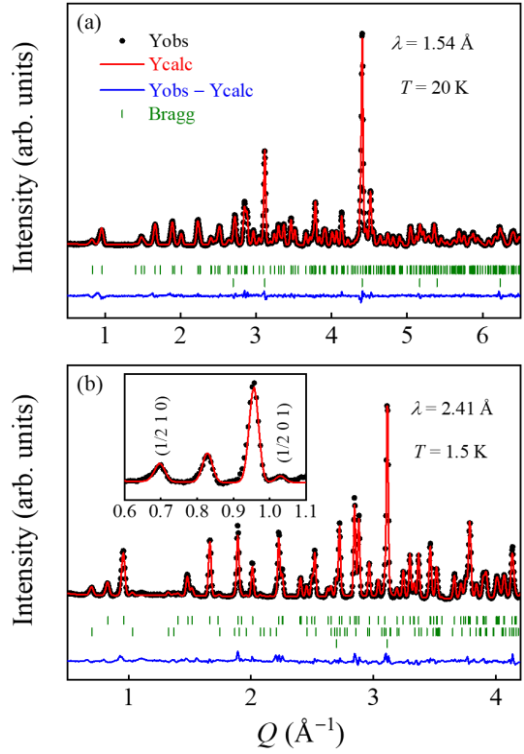

FIG. 3. (a) The NPD data of BaNiTe$_2$O$_7$ and its refinements collected at 20 K with the wavelength of 1.54 Å. (b) shows the data measured at 1.5 K with the wavelength of 2.41 Å. The vertical bars indicate the nuclear and magnetic Bragg peak positions, as well as the aluminum peaks from the sample holder. The inset in (b) shows a zoomed-in view of the low-$Q$ diffraction data that includes the magnetic peaks (1/2,1,0) and (1/2,0,1).

Symmetry-allowed magnetic structure models were created using the irreducible representation analysis with the SARAh software [41] and the magnetic symmetry approach using the MAXMAGN program at the Bilbao Crystallographic Server [42]. The magnetic Ni$^{2+}$ ion was analyzed within the crystal space group *Ama*2 and **k** = (1/2, 1, 0). There is only one possible second-order irreducible representation and, correspondingly, two maximal magnetic space groups $C_am$ and $C_c2$ allowed for the Ni$^{2+}$ ion at the 4$b$ Wyckoff position. The best fit to the data was obtained by adopting a model with the $C_c2$ magnetic space group for a magnetic unit cell defined as $2a \times b \times c$, with respect to the parent lattice. The combined structural and magnetic structural refinement gave an $R$-factor $R_f$ = 6.02 % and $\chi^2$ = 7.85. The refinement result is shown in Fig. 3(b) and the generated magnetic structure is shown in Fig. 4(a) (For the fitting result using the magnetic space group $C_am$ allowed for the Ni$^{2+}$ ion at the 4$b$ Wyckoff position, see Supplemental Material, Fig. S4 [43]). Analogous to BaCoTe$_2$O$_7$ [32], a canted up-up-down-down spin configuration along the zigzag chains has also been found in BaNiTe$_2$O$_7$. The obtained magnetic moment size of Ni$^{2+}$ is 1.8(1)$\mu_B$. The magnetic moment components projected on the three crystallographic axes are $(m_a, m_b, m_c)$ = [-0.4(1), 1.15(5), 1.3(1)]$\mu_B$. The magnetic moments are distributed in a noncollinear structure with the moments linked by a zigzag bond ($J_1$) aligned in an orthogonal pattern, while the nearest moments connected along the $a$ direction ($J_2$) change direction by 180°. A detailed description of the Ni moments arrangement inside the magnetic unit cell is given in Table II.

TABLE II. Atomic positions and magnetic moments arrangement of Ni atoms corresponding to the magnetic space group $C_c2$ and the magnetic unit cell $2a \times b \times c$.

| Ni ($x$ = 0.125, $y$ = 0.61859, $z$ = 0.8586) | $m_x$ = -0.4, $m_y$ = 1.15, $m_z$ = 1.3 ($\mu_B$) |
|---|---|
| 1/8, $y$, $z$ | $m_x$, $m_y$, $m_z$ |
| 3/8, -$y$, $z$ | $m_x$, $m_y$, -$m_z$ |
| 1/8, $y$ + 1/2, $z$ + 1/2 | -$m_x$, -$m_y$, -$m_z$ |
| 3/8, -$y$ + 1/2, $z$ + 1/2 | -$m_x$, -$m_y$, $m_z$ |
| 5/8, $y$, $z$ | -$m_x$, -$m_y$, -$m_z$ |
| 7/8, -$y$, $z$ | -$m_x$, -$m_y$, $m_z$ |
| 5/8, $y$ + 1/2, $z$ + 1/2 | $m_x$, $m_y$, $m_z$ |
| 7/8, -$y$ + 1/2, $z$ + 1/2 | $m_x$, $m_y$, -$m_z$ |

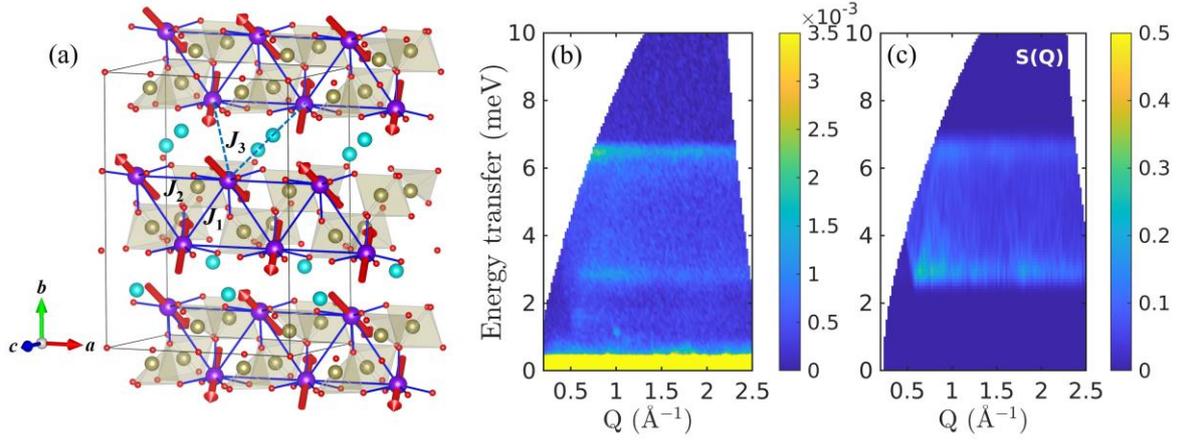

FIG. 4. (a) Magnetic structure of BaNiTe$_2$O$_7$ determined by the neutron-diffraction refinement and verified by DFT calculation. $J_1$, $J_2$, and $J_3$ denote the nearest-, next-nearest-, and next-next-nearest-neighbor exchanges. (b) Inelastic neutron-scattering data obtained from BaNiTe$_2$O$_7$ powder at 1.7 K using the HYSPEC spectrometer. (c) Calculated powder averaged spin-wave excitation spectrum, $S(Q, E)$, using a spin Hamiltonian described in the text.

To investigate the exchange interactions and probe the quasi-one-dimensional nature of BaNiTe$_2$O$_7$, we conducted INS measurements using the HYSPEC spectrometer. Measurements were performed below and above the long-range order transition, at 1.7, 10, and 50 K. As shown in Fig. 4(b), the low energy scattering at 1.7 K is dominated by a band of excitations that extends to approximately 7 meV and with a gap of about 2.5 meV. The scattering intensity decreases as the temperature increases to 10 K and almost vanishes at 50 K (see Fig. S1 in the Supplemental Material [43]). The evolution with the temperature of the inelastic signal confirms its magnetic origin. To describe the observed excitation spectra, we developed a minimal model spin Hamiltonian [see Eq. (1)], which includes two intrachain ($J_1$ and $J_2$) exchange parameters and one interchain ($J_3$) exchange parameter, and a single-ion anisotropy term ($D$) with the direction defined by the unit vector $\hat{e}$ (magnetic easy axis). The vector $\hat{e}$ corresponds to a local symmetry axis of the square pyramid that is perpendicular to a triangular lateral facet, in agreement with the determined magnetic moment direction.

$$\mathcal{H} = \sum_{<i,j>_{n=1,2,3}} J_n S_i S_j + \sum_i D(\hat{e} \cdot S_i)^2 \quad (1)$$

For the nearest-neighbor intrachain couplings, we considered the exchange interactions corresponding to zigzag Ni-Ni bonds of about 4.6 Å ($J_1$) and to the nearest-neighbor bonds along the *a* axis of about 5.56 Å ($J_2$). Both $J_1$ and $J_2$ couplings involve superexchange pathways via Te-O groups, as displayed in Fig. 4(a). The interchain coupling $J_3$ relates to the longer bond of 6.0 Å and is expected to be a much weaker interaction due to the indirect exchange pathways via Ba atoms.

The linear spin-wave theory implemented in the SPINW program was used to test various combinations of $J_n$ and $D$ in an iterative approach. While we cannot uniquely determine the exchange values the closest agreement to the data was obtained with $J_2 \gg J_1$ and $J_3$. The analyses indicate that the main characteristics of the excitation spectrum as well as the magnetic structure can be reproduced using the following parameters: $J_1 = 0.15$ meV, $J_2 = 3.05$ meV, $J_3 = 0.01$ meV, and $D = -0.3$ meV (see Fig. S2 in the Supplemental Material [43]). The positive $J$ values correspond to antiferromagnetic coupling. The calculated powder averaged spectra is shown in Fig. 4(c). We note that the agreement of the model to the data is not fully complete. Notably, there is a stronger intensity at the top of the band region and a possible additional weak excitation mode at very low energies in the experimental data. This indicates an incomplete model under the semi-classical approximation and may point to an intriguing low-dimensional quantum behavior associated with the $S = 1$ Haldane chain.

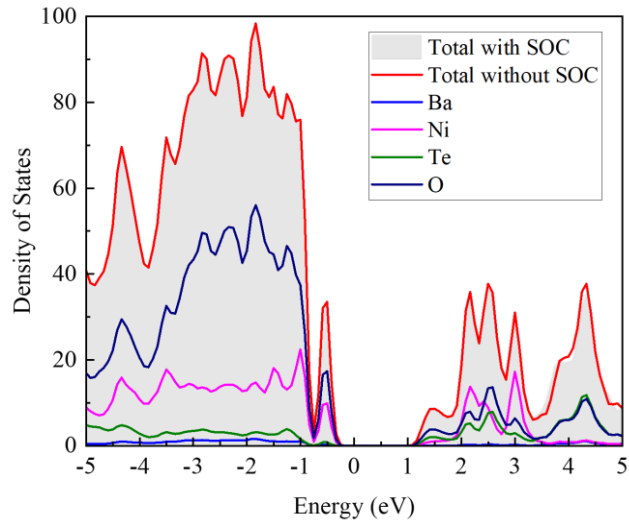

FIG. 5. The density of states of $BaNiTe_2O_7$. The SOC effect is almost negligible to the electronic structure. The hybridization between Ni's 3*d* orbitals and O's 2*p* orbitals contributes to the states

around the Fermi level.

To further verify the magnetic order of BaNiTe$_2$O$_7$, DFT calculation was performed to compare the energies of various AFM orders and ferromagnetic order. It is found that the noncollinear AFM order refined by neutron diffraction is indeed the lowest in energy among all those considered (see Supplemental Material for all configurations and their energies [43]). The optimized orthorhombic lattice constants of canted AFM BaNiTe$_2$O$_7$ are $a$ = 5.561 Å, $b$ = 15.176 Å, and $c$ = 7.259 Å, which are close to the experimental ones. The calculated magnetic moment is ~1.70$\mu_B$/Ni (within the default Wigner-Seitz sphere), as expected from its two half-filled $e_g$ orbitals and also close to the experimental value.

The magnetic orientations from neutron study are also confirmed in our calculation with SOC, leading to ($m_a$, $m_b$, $m_c$) = (-0.417, 1.148, 1.317)$\mu_B$. The exchange coefficients extracted from DFT calculation are $J_1$ = 0.32 meV, $J_2$ = 3.66 meV, and $J_3$ = 0.01 meV. The dominant $J_2$ leads to unambiguous antiparallel spin correlation along the $a$ axis. In fact, the DFT $J_2$ and $J_3$ are close to the values fitted from the inelastic neutron experiment, while DFT $J_1$ also qualitatively agrees with the experimental expectation. Also, magnetic anisotropy is calculated. The easy axis is the $c$ axis, which is -0.34 meV lower than that along other axes. The energy difference is in accordance with the result of experiment.

In quasi-one-dimensional spin chains, the interchain interaction and single-ion anisotropy are two very important factors in defining the boundary of the quantum phase transition between spin-liquid non-magnetic and Ising-like ordered states. A phase diagram for the threshold values of the interchain coupling and anisotropy has been proposed by Sakai and Takahashi [23]. In this context, our system is comparable to the quasi-one-dimensional systems PbNi$_2$V$_2$O$_8$ and SrNi$_2$V$_2$O$_8$ [9,10,26]. Using the experimental values for the intrachain $J$ ~ 9 meV, interchain $J'$ ~ 0.15 meV and $D$ ~ -0.5 meV, these two compounds were positioned just below the spin-liquid Ising transition line, enough to be considered as Haldane-gap antiferromagnets [26].

The density of states of canted AFM BaNiTe$_2$O$_7$ with/without SOC is shown in Fig. 5. The estimated band gap is ~1.2 eV. The valence bands maximum is mainly contributed by the hybridized Ni's 3$d$ and the O's 2$p$ orbitals, while the conduction band minimum is mostly from Ni's upper Hubbard bands of 3$d$ orbitals. As expected, the electronic structure is almost not affected by the SOC effect, which leads to very weak magnetocrystalline anisotropy (i.e., a very small $D$ as mentioned above).

## IV. CONCLUSION

In summary, we investigate the structure and magnetism of BaNiTe$_2$O$_7$ by x-ray diffraction, magnetic susceptibility, specific heat, neutron powder diffraction, and inelastic neutron-scattering measurement techniques. The experimental study was complemented by first-principles DFT calculations. Our results reveal that BaNiTe$_2$O$_7$ crystallizes in a noncentrosymmetric space group *Ama*2 and Ni$^{2+}$ ions form spin chains with a commensurate AFM structure characterized by the propagation vector **k** = (1/2, 1, 0) below $T_N$ ~ 5.4 K. The refined Ni$^{2+}$ magnetic moment is 1.84$\mu_B$ at 1.5 K close to our DFT calculation. Inelastic neutron scattering revealed a strong one-dimensional character of the underlying spin dynamics. In this work, we provide a platform to study the low-dimensional quantum magnetism in close proximity to Haldane chain physics.

**ACKNOWLEDGMENTS**

This work was supported by the National Natural Science Foundation of China (Grant Nos. 12074111, 52272108, 92163210, 11834002). National Natural Science Foundation of China No. 12204160 and the Hubei Provincial Natural Science Foundation of China with Grant No. 2023AFA105. A portion of this research used resources at High Flux Isotope Reactor and Spallation Neutron Source, DOE Office of Science User Facilities operated by the Oak Ridge National Laboratory. Y.M.G. and S.D. thank the Big Data Center of Southeast University for providing the facility support for the calculations.